\documentclass[aps,prb,twocolumn,superscriptaddress,groupedaddress]{revtex4}
\usepackage{graphicx}  
\usepackage{dcolumn}   
\usepackage{bm}        
\usepackage{amssymb}   
\usepackage[version=3]{mhchem} 
\usepackage{dsfont}
\usepackage{url}
\usepackage{subfigure}

\newcommand{\EF}{E_\mathrm{F}} 

\DeclareMathAlphabet      {\mathbfit}{OML}{cmm}{b}{it}

\newcommand{\ket}[1]{\ensuremath{|#1\rangle}}
\newcommand{\bra}[1]{\ensuremath{\langle #1|}}

\usepackage{color}

\usepackage[normalem]{ulem}

\begin{document}

\title{Interplay of screening and superconductivity in low-dimensional materials}

\author{G. Sch\"onhoff}
\affiliation{Institut f\"ur Theoretische Physik, Universit\"at Bremen, Otto-Hahn-Allee 1, 28359 Bremen, Germany}
\affiliation{Bremen Center for Computational Materials Science, Universit\"at Bremen, Am Fallturm 1a, 28359 Bremen, Germany}
\author{M. R\"osner}
\affiliation{Institut f\"ur Theoretische Physik, Universit\"at Bremen, Otto-Hahn-Allee 1, 28359 Bremen, Germany}
\affiliation{Bremen Center for Computational Materials Science, Universit\"at Bremen, Am Fallturm 1a, 28359 Bremen, Germany}
\author{R. E. Groenewald}
\affiliation{Department of Physics and Astronomy, University of Southern California, Los Angeles, CA 90089-0484, USA}
\author{S. Haas}
\affiliation{Department of Physics and Astronomy, University of Southern California, Los Angeles, CA 90089-0484, USA}
\author{T. O. Wehling}
\affiliation{Institut f\"ur Theoretische Physik, Universit\"at Bremen, Otto-Hahn-Allee 1, 28359 Bremen, Germany}
\affiliation{Bremen Center for Computational Materials Science, Universit\"at Bremen, Am Fallturm 1a, 28359 Bremen, Germany}

\date{\today}

\begin{abstract}
A quantitative description of Coulomb interactions is developed for  two-dimensional superconducting materials, enabling us to compare intrinsic with external screening effects, such as those due to substrates. 
Using the example of a doped monolayer of \ce{MoS2} embedded in a  tunable dielectric environment, we demonstrate that the influence of external screening is limited to a length scale, bounded from below by the effective thickness of the quasi two-dimensional material and from above by its intrinsic screening length. 
As a consequence, it is found that unconventional Coulomb driven superconductivity cannot be induced in \ce{MoS2} by tuning the substrate properties alone. 
Our calculations of the retarded Morel-Anderson Coulomb potential $\mu^*$ reveal that the Coulomb interactions, renormalized by the reduced layer thickness and the substrate properties, can shift the onset of the electron-phonon driven superconducting phase in monolayer \ce{MoS2} but do not significantly affect the critical temperature at optimal doping.
\end{abstract}

\maketitle

\section{Introduction}

Various quasi-two-dimensional (2d) materials are known to exhibit a competition of superconducting (SC), charge density wave, and magnetic  phases \cite{somoano_alkaline_1975, woollam_proceedings_1977, paglione_high-temperature_2010,taniguchi_electric-field-induced_2012, ye_superconducting_2012, xi_strongly_2015, yu_gate-tunable_2015, glasbrenner_effect_2015, cao_quality_2015, jo_electrostatically_2015, shi_superconductivity_2015}, with notably different dependences of the resulting phase diagrams on the number of layers.
While in some systems, such as the Fe-based superconductors, the highest transition temperatures $T_\mathrm{c}$ are reached in the monolayer limit (e.g. $T_\mathrm{c}\sim 100\,$K in FeSe on SrTiO$_3$ substrates \cite{yan_interface_2012, coh_large_2015, ge_superconductivity_2015, tang_interface_2016}), several superconducting transition metal dichalcogenides (TMDCs) show exactly the opposite trend of a decreasing $T_\mathrm{c}$ when  monolayer thickness is approached \cite{frindt_superconductivity_1972, xi_strongly_2015, yu_gate-tunable_2015,cao_quality_2015,biscaras_onset_2015, costanzo_gate-induced_2016}. 
These observations point towards several competing effects in layered materials, including enhanced quantum fluctuations,  singularities in the electronic density of states and response functions, strain, tunable Fermi surface topologies etc., all of which potentially contribute to these trends. 
In all these cases an important common factor pertains to how the renormalization of Coulomb interactions due to reduced material dimensionality and environmental screening affects the superconducting transition when approaching the monolayer limit. 

A representative example for such strongly thickness dependent superconductivity is molybdenum disulfide, \ce{MoS2}.
This material becomes superconducting upon electron doping, e.g. via intercalation of alkali atoms \cite{somoano_alkaline_1975, woollam_proceedings_1977} or by gating \cite{taniguchi_electric-field-induced_2012, ye_superconducting_2012, jo_electrostatically_2015, shi_superconductivity_2015}. 
Its temperature-versus-doping phase diagram  is characterized by a dome-shape superconducting region, with critical temperatures on the order of a few K at optimal doping, and by a highly anisotropic response to magnetic fields \cite{lu_evidence_2015,saito_superconductivity_2016}. 
Recent experiments on field effect doped layered \ce{MoS2} have demonstrated superconductivity down to the monolayer limit, where $T_\mathrm{c}$ decreases from $\sim 10$\,K in thicker flakes ($>6$ layers) to $2$\,K for the monolayer \cite{costanzo_gate-induced_2016}. 
The reason behind this evolution remains elusive. 
On the theory side, several mechanisms including purely electronic ones (called \textit{Coulomb driven} hereafter) have been suggested to give rise to superconductivity,  predicting unconventional \cite{roldan_interactions_2013, yuan_triplet_2015} and possibly topologically non-trivial types of superconducting order \cite{Yuan_topol_SC_PRL2014}. 
In contrast,  more conventional pathways to superconducting pairing resulting from electron-phonon coupling have also been proposed \cite{ge_phonon-mediated_2013, rosner_phase_2014,das_superconducting_2015}. 
However, in all of these scenarios it is unclear to which extent the renormalized Coulomb interactions  affect superconductivity when approaching the monolayer limit.

In this paper, we develop a quantitative theory of how Coulomb interactions affect the superconducting transition in \ce{MoS2} as a representative example of TMDCs.
Using ab-initio calculations we derive effective Coulomb coupling constants, where we account for extrinsic and intrinsic screening within the random phase approximation (RPA). 
On this basis, we show that a purely Coulomb driven superconducting phase with an order parameter that has opposite signs in different valleys \cite{roldan_interactions_2013} is not favored in \ce{MoS2}. 
Rather, additional strong renormalization of the interactions, e.g. by spin fluctuations \cite{yuan_triplet_2015}, would be needed to obtain Coulomb driven superconductivity, here.
We find this to be true independently of the dielectric environment of the substrate. 
 For the scenario of phonon mediated SC \cite{ge_phonon-mediated_2013, rosner_phase_2014,das_superconducting_2015}, we show that the phonon mediated electron-electron attraction generally overcomes the Coulomb repulsion when a Lifshitz transition takes place and additional Fermi pockets become available. 
The intrinsic screening of TMDCs at their superconducting transition is shown to be typically so large that it renders external substrate screening rather unimportant for the SC transition despite the atomic scale proximity of the substrate. 
As a consequence, we find that the reduced transition temperatures in monolayer \ce{MoS2} as compared to the bulk are not due to a lack of Coulomb screening in the monolayer limit.

\section{Electronic structure and Coulomb interactions}

\begin{figure}%
		\includegraphics[width=\columnwidth]{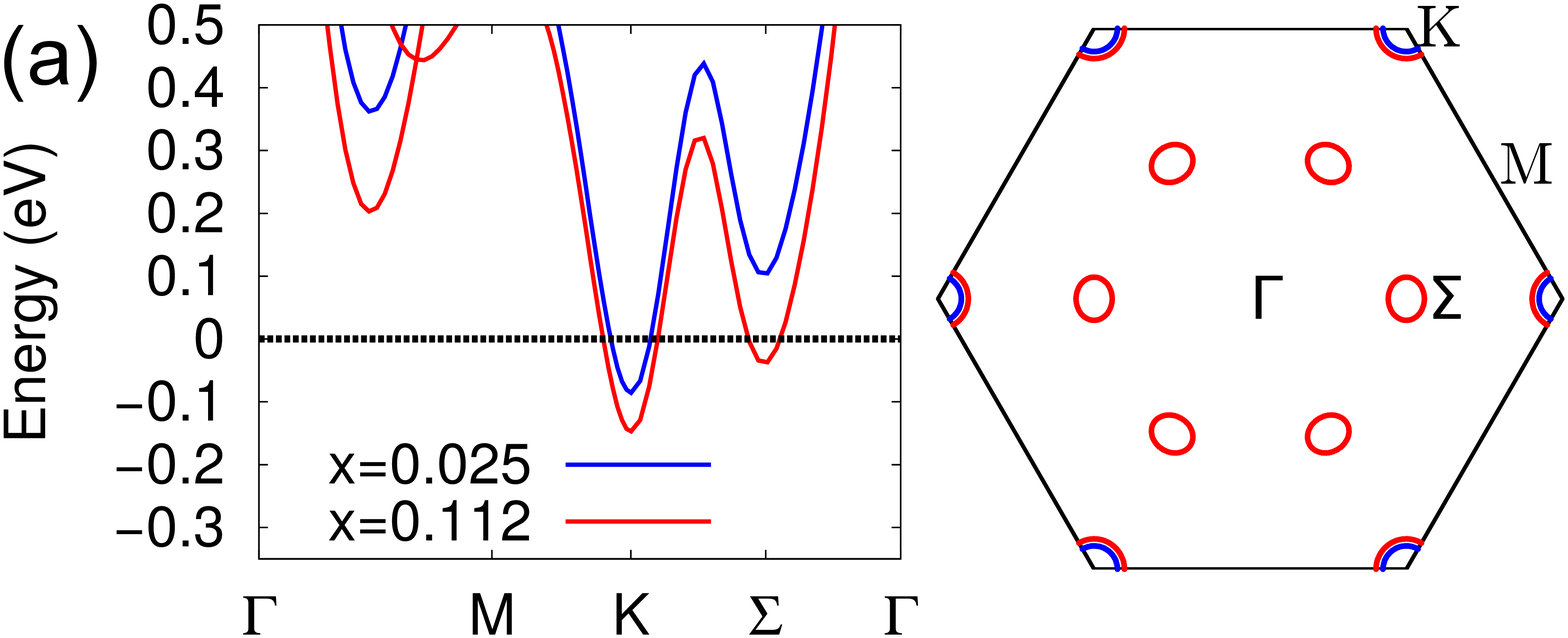}
		\hspace*{-0.5cm}
		\includegraphics[width=\columnwidth]{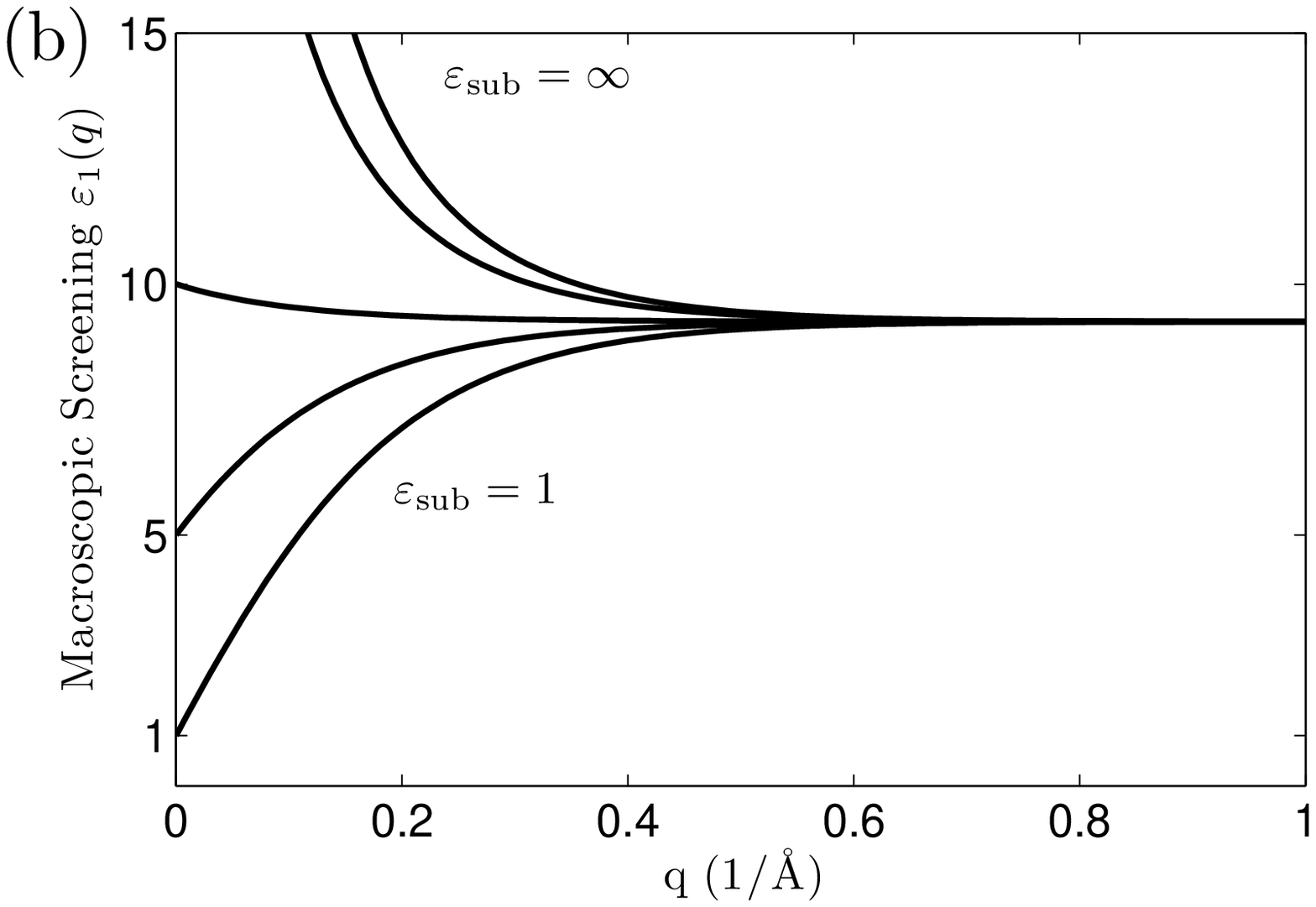}{\hspace*{-0.2cm}\makebox[\wd1][r]{\raisebox{1cm}{\includegraphics[width=0.14\textwidth]{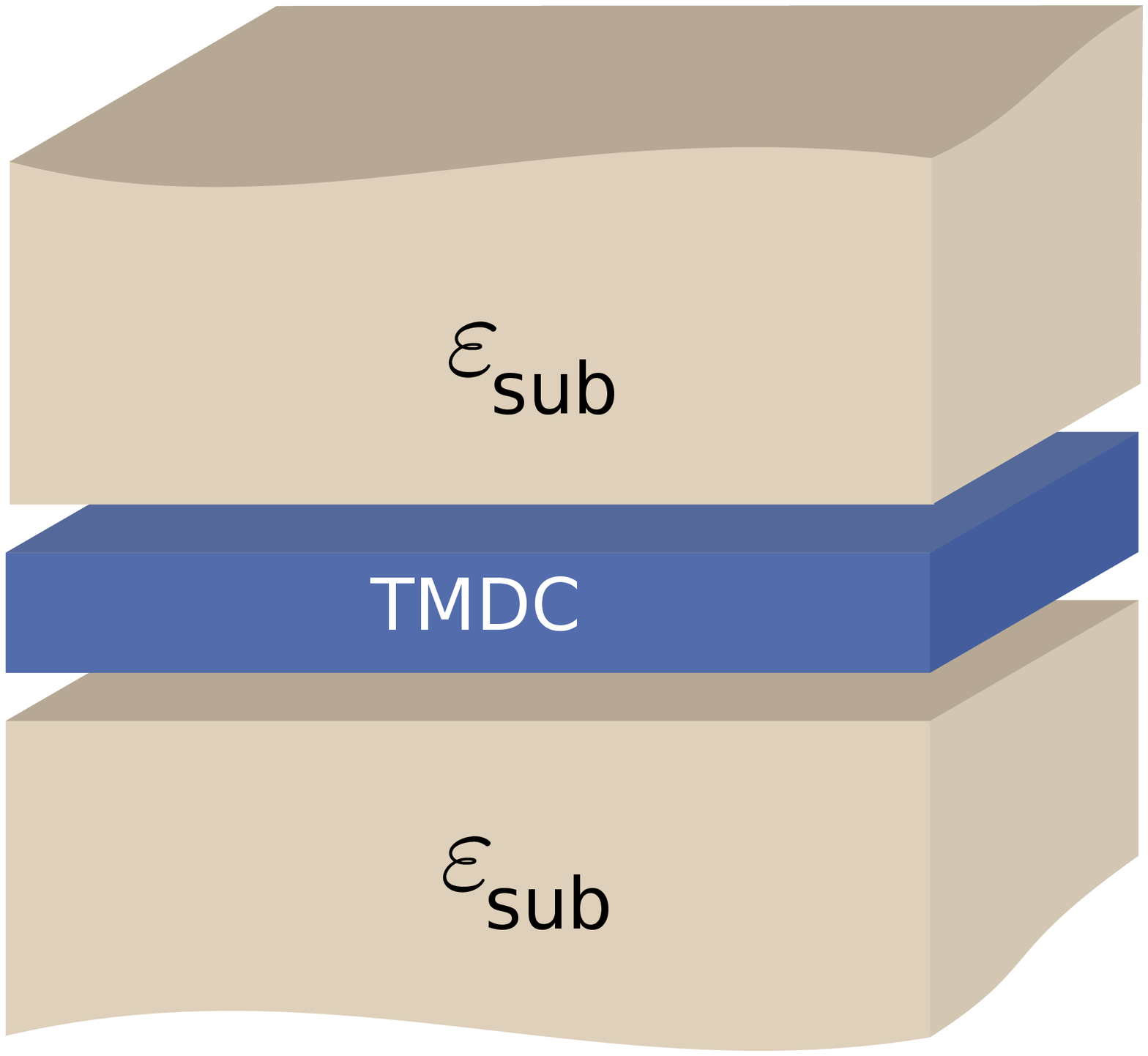}}}}
		\includegraphics[width=\columnwidth]{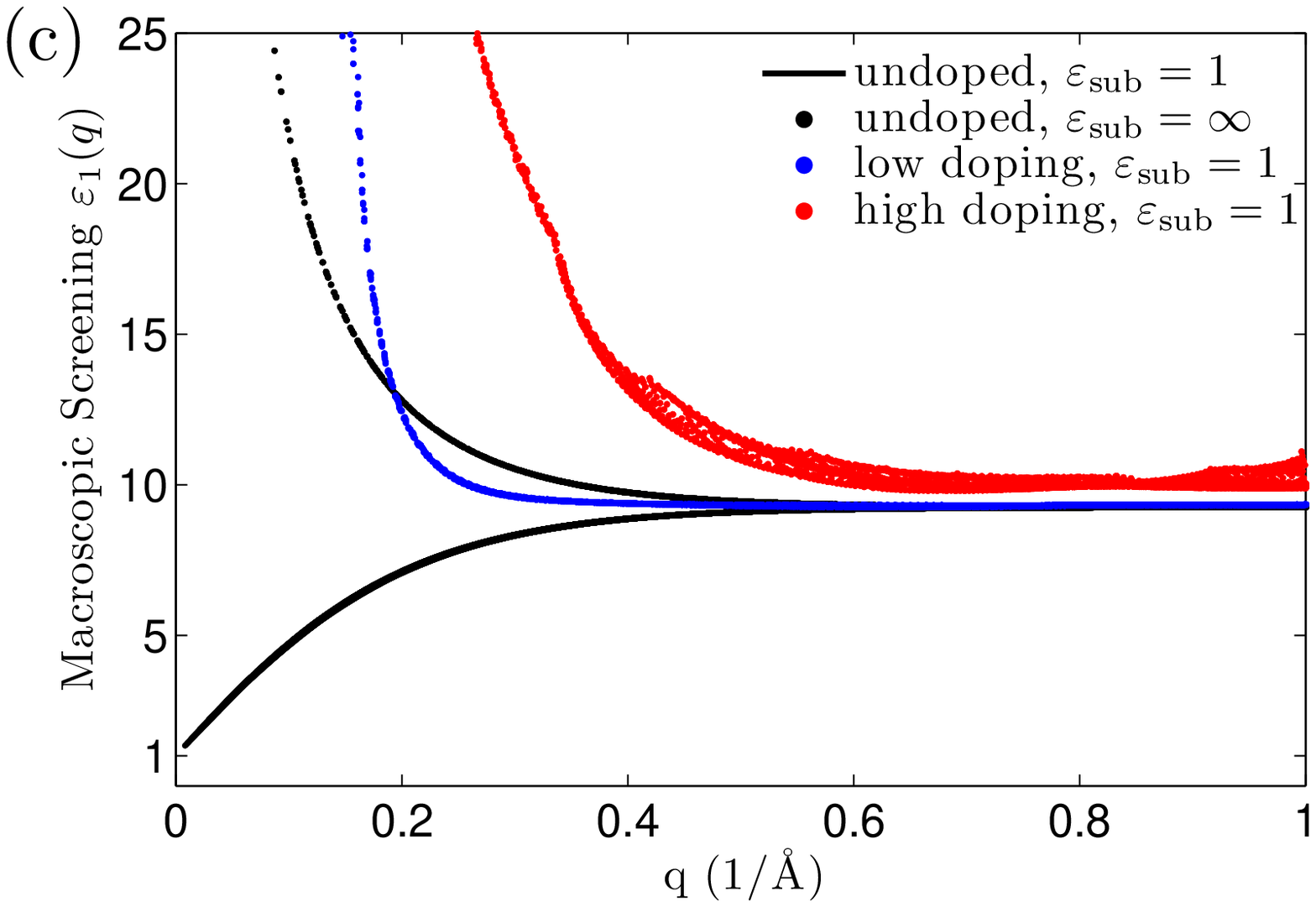}
		\caption{
			(Color online) \textbf{(a)} Band structures and Fermi surfaces of \ce{MoS2} at doping levels $x=0.025$ and $0.112$ electrons per unit cell; the black line is the Fermi energy. \textbf{(b)} Macroscopic screening $\varepsilon_1(q)$ from fit model; the substrate dielectric constants used for $\varepsilon_1(q)$  are $\varepsilon_\mathrm{sub}=\,$1, 5, 10, 50, $\infty$. \textbf{(c)} Full screening for the undoped system with different surroundings as well as for low and high doping with vacuum surrounding.
		}%
		\label{fig:bands_FS}%
	\end{figure}

	Using density functional theory (DFT) as implemented in Quantum Espresso \cite{giannozzi_quantum_2009} we obtain the band structure of monolayer \ce{MoS2}, neglecting the effects of spin-orbit coupling. 
	The resulting Fermi surfaces and the corresponding segments of the band structure are shown in Fig. \ref{fig:bands_FS}(a) for two different electron doping levels. 
There are two prominent minima in the lowest conduction band \cite{kuc_influence_2011}. 
The lower-energy minimum is at the K-points, whereas  the higher-energy minimum  is at $\Sigma= \frac{1}{2} \overline{\Gamma \mathrm{K}}$. 
Hence, this multiple-valley band structure of \ce{MoS2} represents a situation where the Fermi surface topology changes with electron doping. 
At sufficiently low doping ($x \lesssim 0.07$, $x$ in electrons per unit cell), the Fermi pockets are all centered around the K-points in the Brillouin zone corners [blue lines in right panel of Fig. \ref{fig:bands_FS}(a)], whereas for $x\gtrsim0.07$ the conduction band minima at $\Sigma$ are also populated by electrons [red lines in right panel of Fig. \ref{fig:bands_FS}(a)]. 
	As we show below,  such a Lifshitz transition  has a profound influence on the competition between Coulomb repulsion and electron-phonon coupling.
	We refer to doping levels of $x<0.07$ as \textit{low doping} in the following, whereas \textit{high doping} is used to describe situations where six additional Fermi sheets around $\Sigma$ exist, $x>0.07$. 
	
		In the following we specifically examine the influence of  screened Coulomb interactions on superconducting pairing, which depends on the electron doping level and on the dielectric environment. 
The available \emph{screening channels} can be divided into \emph{internal} and \emph{external} channels, which both contribute to the strength of the renormalized Coulomb interactions. 
Here, internal processes refer to the screening due to transitions between electronic states within the \ce{MoS2} layer.
The external screening arises due to the polarizability of adjacent substrates or adsorbates with dielectric constant $\varepsilon_\text{sub}$. 

In order to obtain Coulomb matrix elements in the Wannier basis we start with the construction of an effective Wannier-Hamiltonian \cite{mostofi_updated_2014} from projections onto the three dominant $d$-orbitals $\alpha$, $\beta \in \mathrm{\{ d_{z^2}, d_{xy}, d_{x^2y^2} \} }$ of the Mo atoms as described in more detail in Ref. [\onlinecite{groenewald_valley_2016}]. 
Doping is modeled as a rigid shift of the Fermi energy in the undoped band structure.
The resulting renormalized Coulomb coupling constants are obtained in the following way (see the appendix A for more details):
	First, we derive realistic screened interaction matrix elements in the Wannier basis $\alpha$, $\beta$ for the freestanding undoped material via RPA calculations using the Spex and Fleur software codes \cite{friedrich_efficient_2010, FLEUR}. 
	The bare $U_{\alpha \beta}(q)$ and intrinsically screened (undoped material with \emph{inter-band} transitions only) matrix elements $V_{\alpha \beta}(q)$ are then parametrized as functions of momentum transfer $q$. 
	Next, the \emph{external} screening effects are accounted for by solving  the Poisson equation for a continuous medium representing the dielectric environment.
	Specifically, we consider the geometric  substrate-monolayer-substrate arrangement depicted in the inset of Fig. \ref{fig:bands_FS}(b). 
	This allows us to compute the screened matrix elements $V_{\alpha \beta}(q)$ using the recently developed Wannier function continuum electrostatics approach (WFCE) \cite{rosner_wannier_2015}. 
To this end, we need to assign a physical thickness $d\approx9.1\,$\AA\ to the monolayer of \ce{MoS2}.

	All screening processes are redered by the dielectric function $\varepsilon(q)$, which is actually a dielectric matrix (see appendix A). 
	In Fig.\,\ref{fig:bands_FS}(b) we show the macroscopic $\varepsilon_1(q)$ as resulting from inter-band and external screening for different dielectric environments of the \ce{MoS2} monolayer \footnote{The function $\varepsilon_1(q)$ describes macroscopic screening effects. 
	The functions $\varepsilon_{2,3}$, describing only microscopic screening, are discussed in the appendix.}. 
	In the long-wavelength limit, $\varepsilon_1(q)$ is fully determined by the dielectric background:  $\varepsilon_1(q \to 0)=\varepsilon_\text{sub}$. 
	In the opposite limit ($q \gtrsim 1\,$\AA$^{-1}$), screening in the undoped case is solely due to the microscopic inter-band polarizability of the \ce{MoS2} layer itself, $\varepsilon_1(q)\approx 9.3\equiv\varepsilon_\infty$, but unaffected by the dielectric environment \cite{andersen_dielectric_2015, rosner_wannier_2015}.

	In addition to the external and inter-band screening, we also have to include metallic intra-band screening by the conduction electrons in the case of electron doping. We then arrive at the fully screened static Coulomb interaction $\hat W(q)= \hat V(q) \, \left(1-\hat V(q)\hat\Pi_0(q)\right)^{-1}$, where $\hat \Pi_0(q)$ is the intra-band polarizability and where $\hat W(q)$, $\hat V(q)$ and $\hat \Pi_0(q)$ are   matrices in the Wannier function basis. 
	The polarizability is obtained using RPA for the lowest conduction band \cite{groenewald_valley_2016}. 
	
	In Fig. \ref{fig:bands_FS}(c) we compare the full dielectric functions for the undoped system in different dielectric surroundings (i.e. $\varepsilon_\text{sub} = 1$ and $\varepsilon_\text{sub} = \infty$) to the free standing ($\varepsilon_{\rm sub}=1$) doped system at low ($x\approx0.02$) and high ($x\approx 0.13$) electron doping.
	In all metallic cases, either due to a metallic environment ($\varepsilon_\text{sub} = \infty$) or electron doping of the \ce{MoS2} monolayer, we find a divergent $\varepsilon\sim 1/q$ for small momenta $q$. Furthermore, in the doping induced metallic regime we observe strong dependencies on the doping level.
	
\section{Effects of internal and external screening on Coulomb interactions} 

	Using the fully screened interaction matrix $\hat W(q)$, we can compute the full Coulomb coupling constant
	\begin{equation}
		\label{eq:mu}
		\mu = \frac{1}{N(\EF)} {\sum_{\mathbf{k k'}}} W_{\mathbf{k}\mathbf{k'}} \delta(\epsilon_\mathbf{k}-\EF) \delta(\epsilon_\mathbf{k'}-\EF),
	\end{equation}
	which is the Fermi surface average of the screened Coulomb interaction, including scattering processes with initial states $\{\mathbf{k},\mathbf{k'}\}$ and final states $\{\mathbf{k'},\mathbf{k}\}$ (i.e. $\mathbf{q} = \mathbf{k} - \mathbf{k'}$). 
	In Eq. (\ref{eq:mu}) $\hat W(q) \to W_{\mathbf{k}\mathbf{k'}}=\bra{(\mathbf{k'},\mathbf{k})}\hat W(k-k')\ket{(\mathbf{k},\mathbf{k'})}$ has been transformed from the orbital basis to the band basis, and we only consider the lowest conduction band for $\hat W(q)$ since it is the only band that crosses the Fermi level for the electron doping concentrations considered here. 
	
\begin{figure}[hbt]
	
		\hspace*{-0.7cm}
		\includegraphics[width=0.5\textwidth]{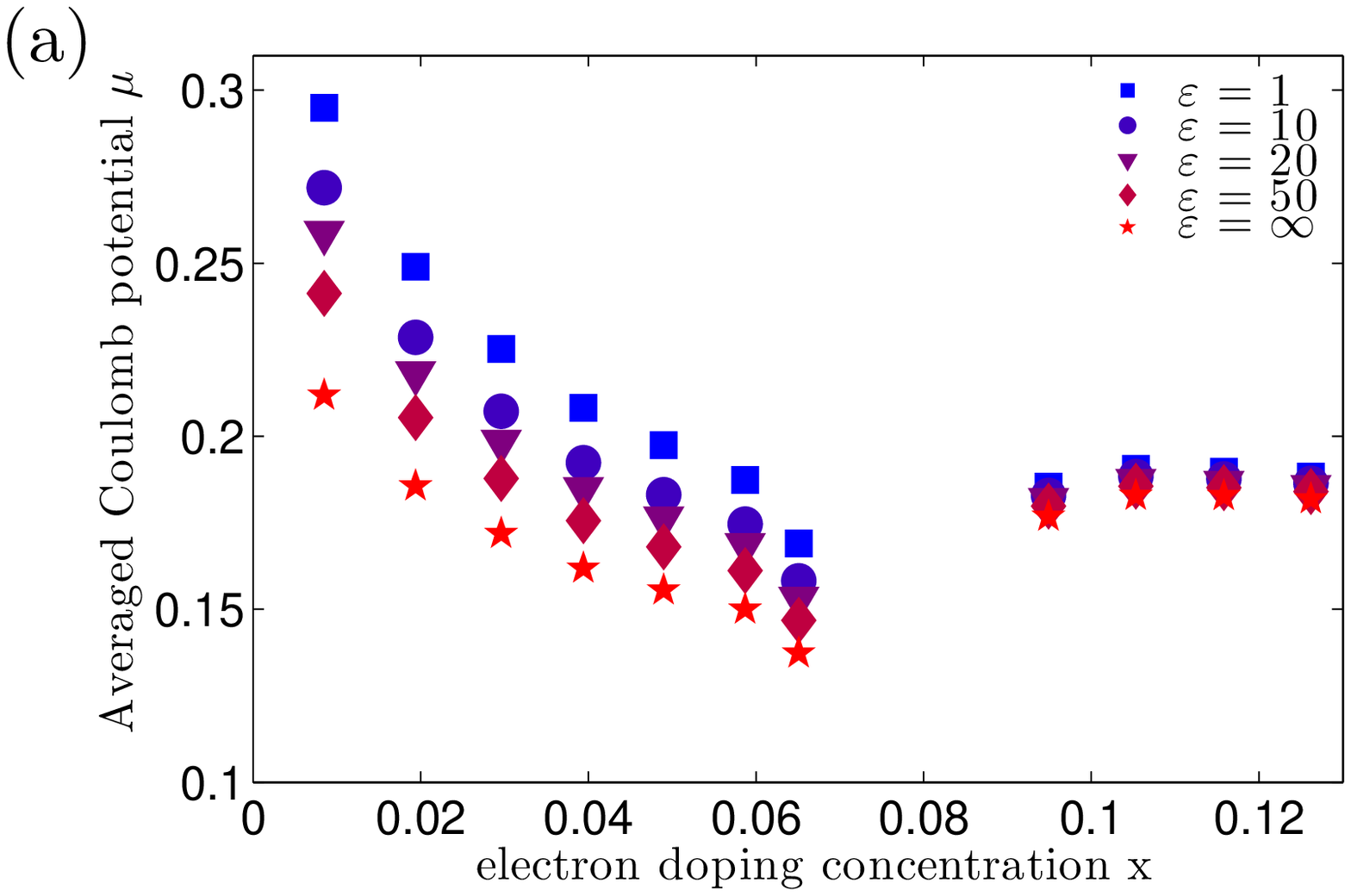}
		
		\hspace*{-0.48cm}
		\includegraphics[width=0.5\textwidth,trim={0 0 1.2cm 0},clip]{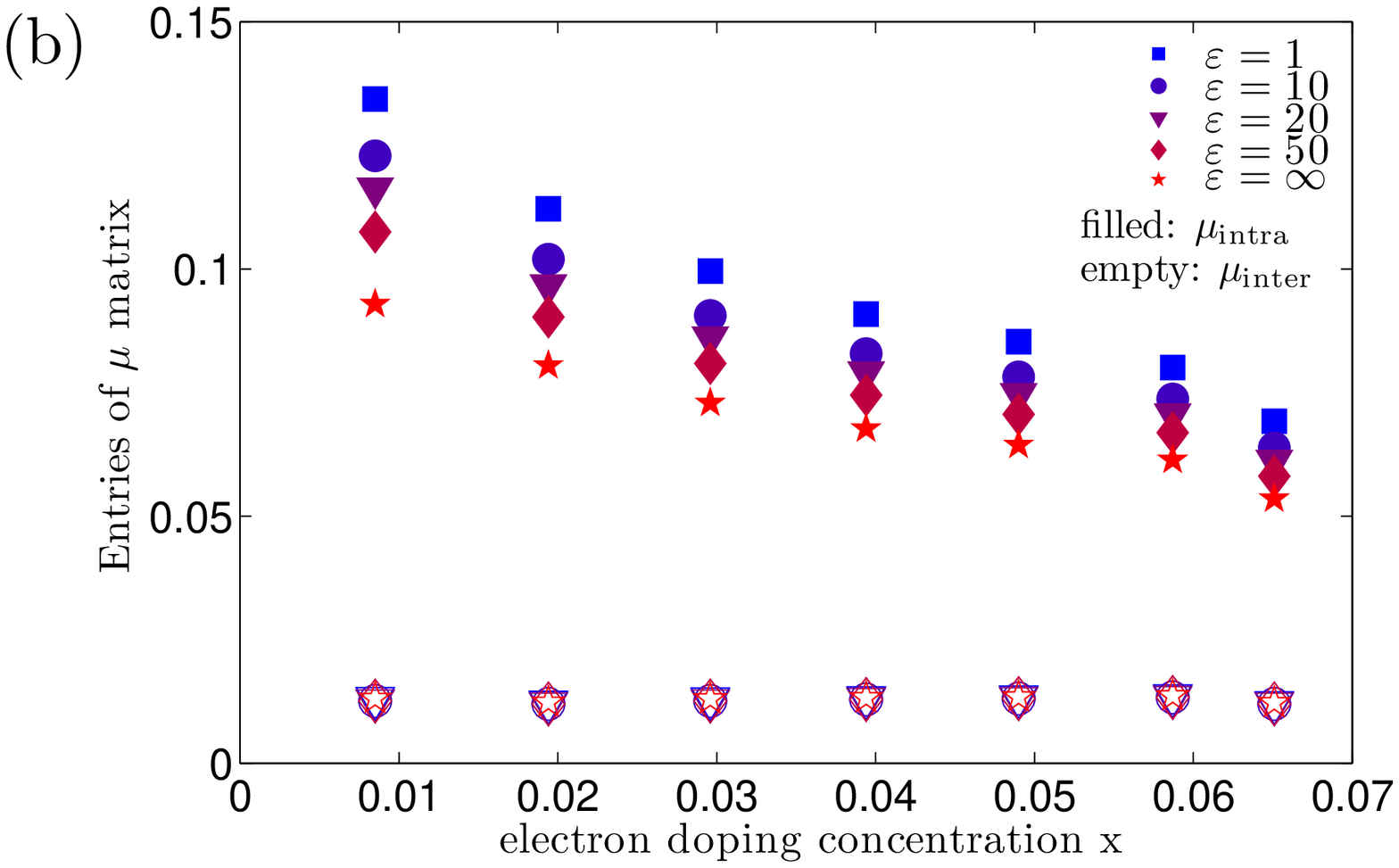}
		
		\caption{
			(Color online) Dependence of Coulomb coupling constants on the electron doping concentration $x$, subject to various $\varepsilon_\text{sub}$ of the dielectric environment. In \textbf{(a)} the full coupling constant $\mu$ is shown. In \textbf{(b)} we present the values of the $2 \times 2$-matrix $\boldsymbol{\mu}$ (cf. Eq. \ref{eq:muinterintra}) for low doping concentrations.
		}
		\label{fig:Mu}
		
	\end{figure}

	The resulting effective Coulomb coupling constants $\mu$ are shown in Fig. \ref{fig:Mu}(a) for different dielectric environments of the \ce{MoS2} layer and in dependence of the electron doping concentration. 
	In the low-doping regime,  $x \lesssim 0.07$, where only two Fermi pockets around K and $\mathrm{K}'$ are present, the coupling $\mu$ is renormalized by up to $\sim 30\%$ via external screening. 
	In contrast, at higher doping concentrations $\mu$ is clearly much less sensitive to its dielectric environment, and variations of $\mu$ due to external screening are limited to $\sim 10\%$. 

	If we account for the multi-valley structure of the Fermi surface [see Fig. \ref{fig:bands_FS}(a)], $\mu$ is no longer a simple scalar but becomes a matrix in the electronic valleys.
	To further investigate the effect of the dielectric environment, we discuss this matrix structure of $\boldsymbol{\mu}$.
	For low doping concentrations, where only the two valleys around the K points are occupied, we obtain the following structure
	\begin{equation}
	\boldsymbol{\mu}_\mathrm{low} = \left(\begin{array}{cc}
			\mu_\mathrm{intra} & \mu_\mathrm{inter} \\
			\mu_\mathrm{inter} & \mu_\mathrm{intra}
			\end{array} \right),
			\label{eq:muinterintra}
	\end{equation}
	where the states $\{\mathbf{k},\mathbf{k'}\}$ in Eq. (\ref{eq:mu}) are in the same valley for $\mu_\mathrm{intra}$ while they are in different valleys for $\mu_\mathrm{inter}$.
	The sum of all matrix elements yields the total coupling constant $\mu$.
	A comparison of external screening effects on intra- and inter-valley Coulomb scattering [Fig. \ref{fig:Mu}(b)] shows that essentially only the intra-valley scattering is affected by the dielectric environment. 
	These observations can be explained intuitively. 
	External screening is most effective when the separation ($\sim 1/q$) of the interacting charges inside the monolayer is \textit{larger} than the distance $\sim \frac{1}{2} d$ to their image charges in the environment but \textit{smaller} than the internal Thomas-Fermi screening length $1/q_{\rm TF}$, i.e. for $q_{\rm TF}<q<\frac{2}{d}$ [cf. Fig. \ref{fig:bands_FS}(c)].
	As a consequence, the influence of the substrate  weakens as soon as $q_{\rm TF}\gtrsim\frac{2}{d}$. 
	Using the effective thickness of $d \approx 9.1\,$\AA, a Thomas-Fermi wave vector $q_{\rm TF} = 2 \pi \mathrm{e}^2 N(\EF) / (A \, \varepsilon_1(q_{\rm TF}))$, and a background dielectric constant on the order of  $\varepsilon_1(q) \approx  \varepsilon_\infty=9.3$ for $q>\frac{2}{d}$, we find that the substrate influence is minor as soon as the density of states at the Fermi level exceeds $N(\EF)\gg0.19$\,/eV per unit cell.

	In \ce{MoS2}, we have $N(\EF) \approx 0.4\,$eV$^{-1}$ and $N(\EF) \approx 2\,$eV$^{-1}$ for low ($x<0.07$) and high electron doping concentrations ($x>0.07$), respectively, which means that substrate influence is weak, especially in the regime of high doping concentrations. 
	However, for sufficiently low doping concentrations, the scattering inside the same K- or ${\rm K}'$-valley can be controlled via the substrates.

\section{Coulomb driven superconductivity}

In general, superconductivity occurs when the total coupling between electrons is attractive: $\mu_\mathrm{tot}<0$.
For conventional phonon-mediated superconductivity, this is the case when the effective coupling between electrons mediated by phonons overcomes the electron-electron repulsion.
Other than that, a superconducting instability is also possible for a purely repulsive interaction if at least one eigenvalue of the coupling matrix $\boldsymbol{\mu}$ is negative, which can be seen from the conditional equation for solutions to the anisotropic BCS equations

\begin{equation}
	\det\left[\mathbf{N} + N(\EF)\, F \, \boldsymbol{\mu} \right] = 0,
\end{equation}
where $\mathbf{N}$ is the diagonal matrix containing the densities of states per valley and $F$ is a function depending on energies, as derived and discussed in appendix B. 
At low doping, i.e. when only two Fermi surface sheets around K and ${\rm K}'$ exist and $\boldsymbol{\mu}$ is a $2 \times 2$-matrix, see Eq. (\ref{eq:muinterintra}), $\mu_{\rm inter} > \mu_{\rm intra}$ would lead to a negative eigenvalue of $\mu$ and thus to a superconducting phase which would be purely electronically mediated with an unconventional sign changing order parameter ($\Delta_\mathrm{K}=-\Delta_\mathrm{K'}$); this case was discussed in Ref. [\onlinecite{roldan_interactions_2013}].
	
\begin{figure}[hbt]
				
				\hspace*{-0.5cm}
				\includegraphics[width=0.5\textwidth]{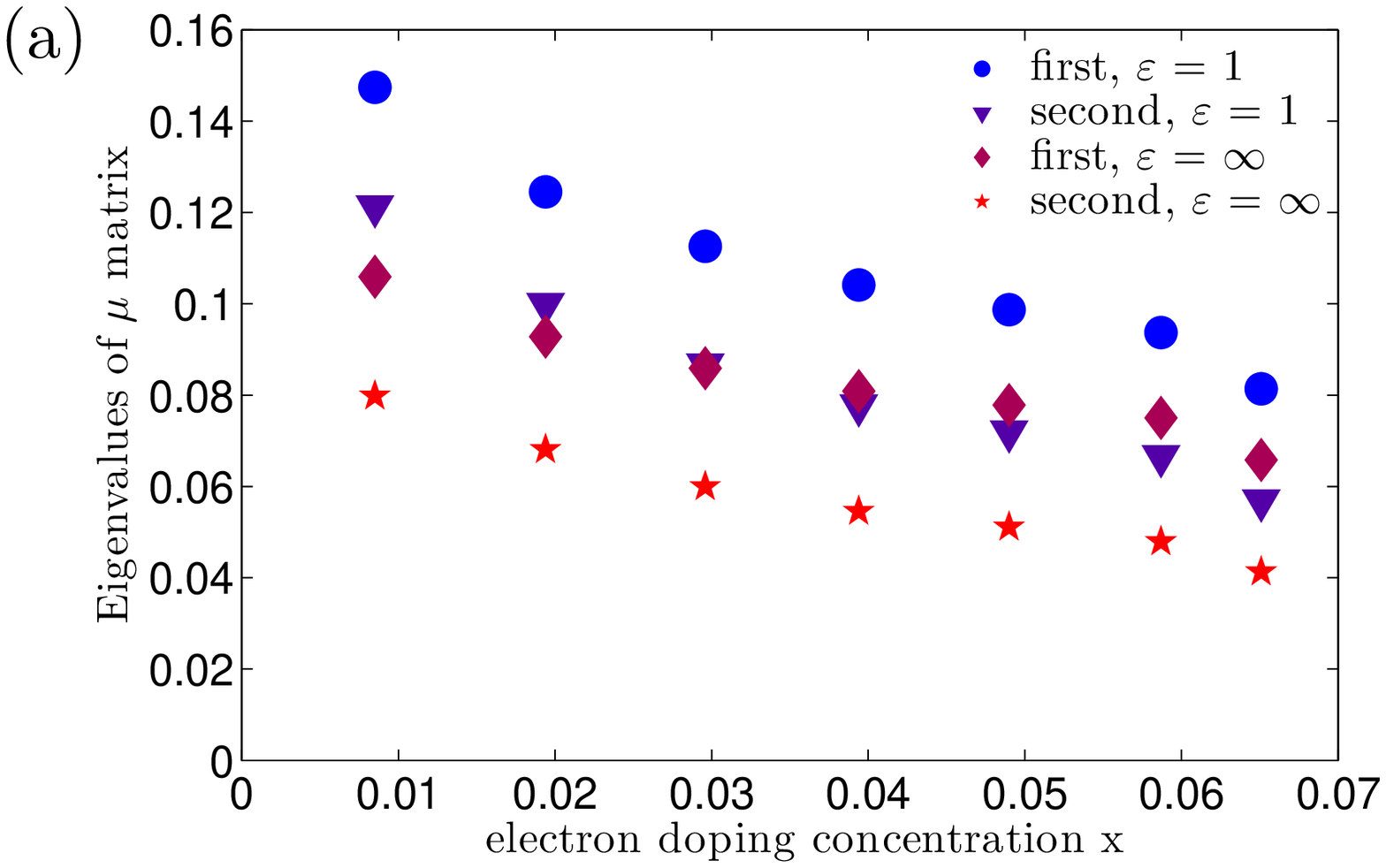}
				
				\hspace*{-0.45cm}
				\includegraphics[width=0.265\textwidth]{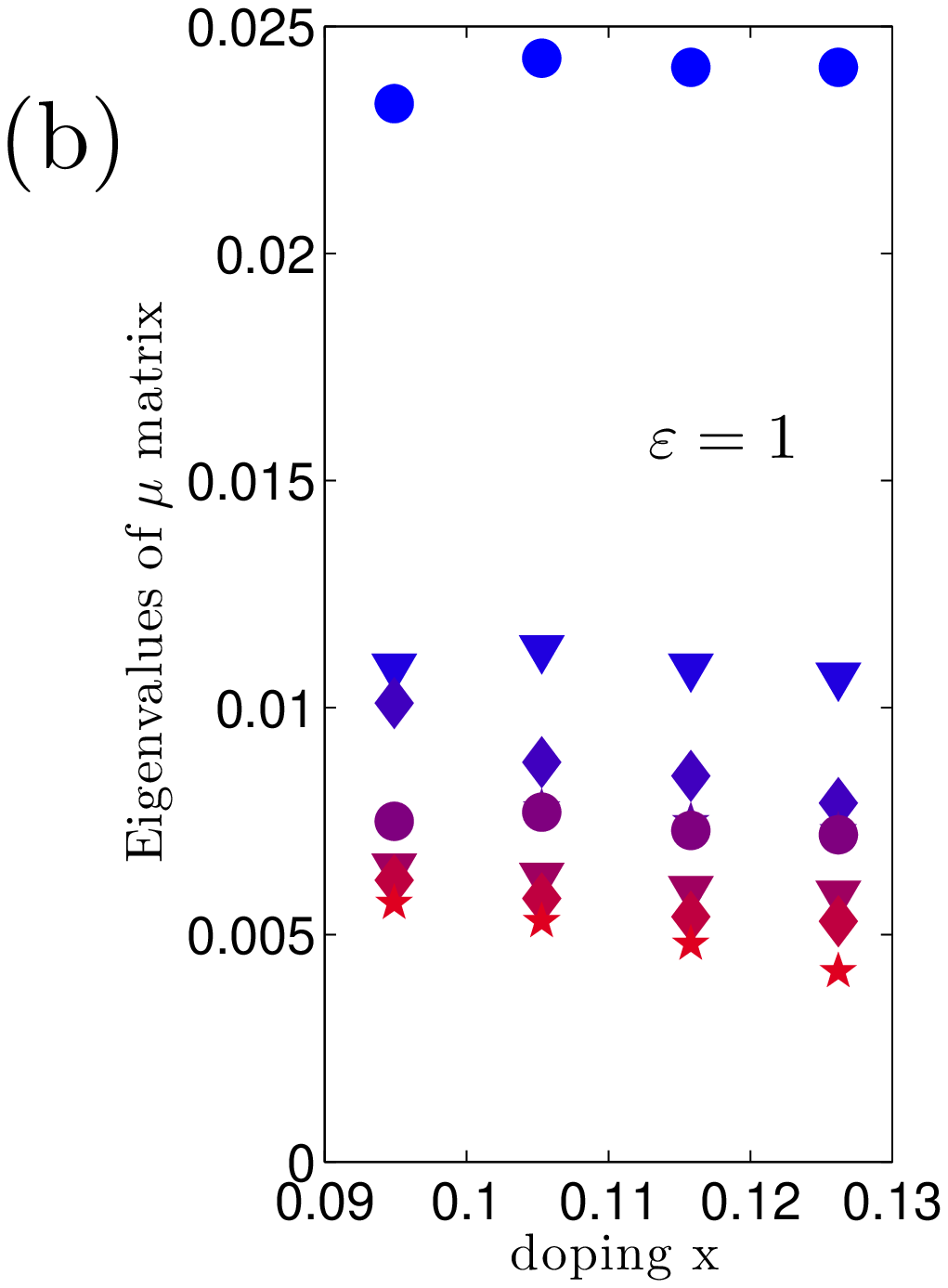}
				\includegraphics[width=0.23\textwidth]{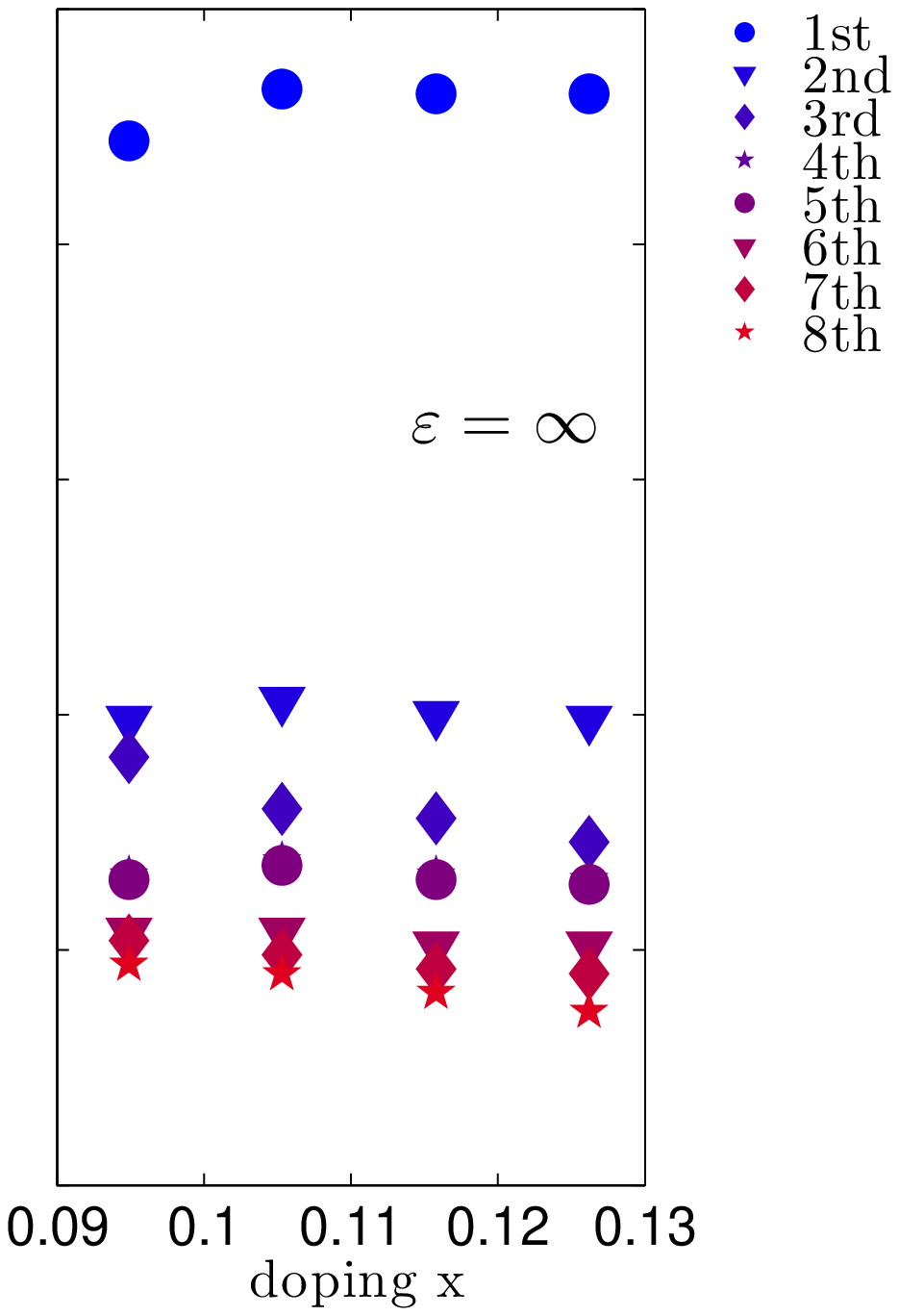}
				
				\caption{
					(Color online) Dependence of the matrix $\boldsymbol{\mu}$ on the electron doping concentration $x$. In \textbf{(a)} we present the eigenvalues of the $2 \times 2$-matrix $\boldsymbol{\mu}$ in the case of low doping for vacuum ($\varepsilon=1$) and metallic ($\varepsilon=\infty)$ surrounding; in \textbf{(b)} the eigenvalues of the $8 \times 8$-matrix $\boldsymbol{\mu}$ in the case of high doping with background dielectric constant $\varepsilon=1$ (left) and $\varepsilon=\infty$ (right) are shown.
				}
				\label{fig:MuEigenValues}
				
			\end{figure}

	However, as one can see from Fig.\,\ref{fig:Mu}(b) and Fig. \ref{fig:MuEigenValues}(a), the intra-valley coupling is always larger than the inter-valley coupling and $\boldsymbol{\mu}$ has only positive eigenvalues, meaning that the described situation is neither realized in freestanding \ce{MoS2} nor can it be achieved using substrates or capping layers with arbitrarily large ($q$ independent) dielectric constants. 
	Furthermore, the same argument holds in the high doping regime where we do not find any negative eigenvalues of the more complex $8 \times 8$-matrix $\boldsymbol{\mu}$, either, see Fig. \ref{fig:MuEigenValues}(b).
	Thus, Coulomb driven superconductivity is not possible in \ce{MoS2} below and above the Lifshitz transition involving the mechanism discussed here.
	
	We conclude that for unconventional electron driven superconductivity in \ce{MoS2} one would need more complex mechanisms involving a stronger renormalization of the interactions at low energies than what can be achieved via substrates \cite{yuan_triplet_2015}.

\section{Electron-phonon coupling driven superconductivity}

	In the framework of Eliashberg theory \cite{Eliashberg}, the Allen-Dynes formula \cite{allen_transition_1975} yields an estimate of the critical temperature, 
	\begin{equation}
		\label{eq:AllenDynes}
		T_\mathrm{c} = \frac{\hbar \omega_\mathrm{log}}{1.2 k_\mathrm{B}} \exp \left[\frac{-1.04 (1+\lambda)}{\lambda (1-0.62\mu^*) - \mu^*}	 			\right],
	\end{equation}
	which accounts for the competition of the phonon driven attractive interaction (entering via the effective coupling strength $\lambda$ and the typical phonon frequency $\omega_\mathrm{log}$) with the repulsive Coulomb interaction expressed by the Morel-Anderson parameter $\mu^*$ \cite{morel_calculation_1962}. 
	The phononic parameters for \ce{MoS2} have been calculated in Refs. [\onlinecite{ge_phonon-mediated_2013, rosner_phase_2014}]. 
	The coefficient $\mu^*$ that describes the Coulomb repulsion is obtained using the formula given by Morel and Anderson \cite{morel_calculation_1962} for the retarded Coulomb potential,
	\begin{equation}
		\label{eq:mustar}
		\mu^* = \frac{\mu}{1+\mu \ln[\frac{\EF}{\omega_\mathrm{log}}]}.
	\end{equation}
	
\begin{figure}
			
			\hspace{-0.6cm}
			\includegraphics[width=0.46\textwidth]{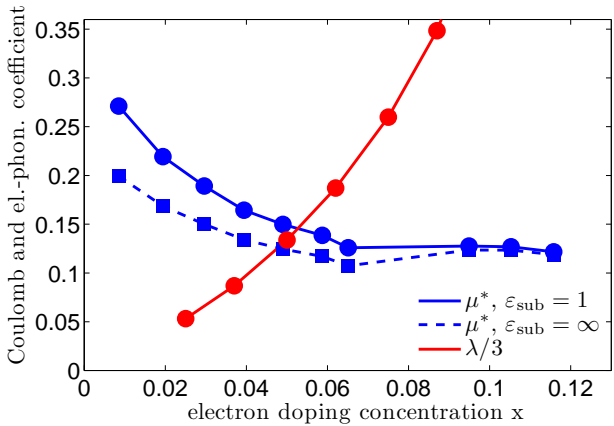}
			
			\caption{
				(Color online) Retarded Coulomb potential $\mu^*$  used in Eliashberg theory and effective electron-phonon coupling strength $\lambda$ in dependence of the electron doping concentration $x$. $\mu^*$  for the freestanding monolayer ($\varepsilon_\mathrm{sub} = 1$) and in the presence of a metallic environment ($\varepsilon_\mathrm{sub} = \infty$). $\lambda$ is scaled by a factor of $1/3$.
			}
			\label{fig:MuStar}
			
		\end{figure}

	In Fig. \ref{fig:MuStar}, we  plot the dependence of $\mu^*$ on the electron doping level for freestanding \ce{MoS2} and \ce{MoS2} embedded in a perfect metallic environment. 
	For free-standing \ce{MoS2} we observe a decrease of $\mu^*$ from $\mu^* > 0.25$ to $\mu^* \lesssim 0.15$ for $x \lesssim 0.07$, which is caused by the corresponding decrease in $\mu$ and the decrease in the phonon frequency $\omega_\mathrm{log}$ (see Ref. [\onlinecite{rosner_phase_2014}]). 
	At larger electron doping concentrations, $\mu$ is basically constant with $\mu^* \sim 0.13$. 
	For \ce{MoS2} embedded in a metallic environment, $\mu^*$ shows essentially the same trend with the only difference in comparison to the free-standing case being a slight reduction of $\mu^*$, particularly at low doping, which could shift the onset of the superconducting phase to a lower doping concentration than the critical concentration for the freestanding layer.	
		
	A significant $T_\mathrm{c}$ is only reached when the exponent in Eq.\,(\ref{eq:AllenDynes}) is close to $-1$ or larger, especially when the electrons mainly couple to acoustic phonons and thus $\omega_\mathrm{log}$ is rather small, e.g. $T_\mathrm{c} \gtrsim  \frac{\hbar \omega_\mathrm{log}}{1.2 k_\mathrm{B}} e^{-2}$. 
	To achieve this, $\lambda>3\,\mu^*$ has to be realized for the range of $0.1 < \mu^* < 0.3$ found here. 
	From the comparison of $\mu^*$ and $\lambda/3$ in Fig. \ref{fig:MuStar}, we see that a significant $T_\mathrm{c}$ (as ocurring for $\lambda > 3 \,\mu^*$) can only be observed once $x\gtrsim 0.07$, i.e. when both valleys in the conduction band are occupied by electrons.

	We thus conclude that the frequent use of a constant for the Coulomb pseudopotential, e.g. $\mu^*=0.13$ \cite{mcmillan_transition_1968, ge_phonon-mediated_2013, rosner_wannier_2015}, is not sufficient in the case of electron doped \ce{MoS2} to describe the influence of the Coulomb interaction directly at the transition to the superconducting phase. 
	However, the drop in the critical temperature of TMDCs \cite{frindt_superconductivity_1972, xi_strongly_2015, yu_gate-tunable_2015,cao_quality_2015, costanzo_gate-induced_2016} when going from the bulk or multilayer-system to a monolayer cannot be caused by enhanced Coulomb interactions, because the values of the electron-phonon coupling are much larger than the $\mu^*\approx0.13$, which we find in the region of optimal doping independently of the dielectric environment of the \ce{MoS2} monolayer.
	
	It is important to note that the present work does not include the possible effects of disorder on the phase diagram.
	These effects can be very significant and can lead to a reduction of the critical temperature of the superconducting phase, especially in systems with low dimensionality. \cite{finkelstein_suppression_1994}
	However, disorder effects strongly depend on the experimental situation and the preparation of the sample which is clearly out of the scope of this paper.

\section{Conclusions}

	The microscopic description of the Coulomb interactions in the electron doped monolayer \ce{MoS2} developed here reveals a clear decrease of the retarded Coulomb potential with increasing doping, which renders the frequent use of a constant, doping and material independent $\mu^*$ questionable. 
	Comparing the values for the electron-phonon interaction in Ref. [\onlinecite{rosner_phase_2014}] with the retarded Coulomb potential $\mu^*$ and the valley decomposed electron-electron interaction coupling constants presented here, we conclude that the superconductivity in \ce{MoS2} is electron-phonon driven and has its onset at electron doping levels for which both valleys in the conduction band are occupied. 
	The effects of substrates turn out to be relatively small, at least around optimal doping, and we find that the experimentally observed reduction of the critical temperature upon approaching the monolayer limit \cite{costanzo_gate-induced_2016} is not caused by enhanced Coulomb interactions, i.e. lack of screening as the dimensionality of the material is reduced. 
	This conclusion should be generally applicable also to other superconducting 2d materials such as \ce{NbSe2} \cite{cao_quality_2015} and particularly the electron doped TMDCs like \ce{WS2} or \ce{MoSe2} \cite{jo_electrostatically_2015, shi_superconductivity_2015} because of their similar electronic and phononic structure.

\textit{Acknowledgments}:
	S.H. would like to the thank the Humboldt Foundation for support. This work was supported by the European Graphene Flagship and by the Department of Energy under Grant No. DE-FG02-05ER46240. The numerical computations were carried out on the Norddeutscher Verbund zur F\"orderung des Hoch- und H\"ochstleistungsrechnens (HLRN) cluster.

 \begin{appendix}	
	\section{Parametrization of realistic screening and Coulomb interaction matrix elements}
	Parts of our calculations on and parametrization of the Coulomb interaction were previously described in Ref. [\onlinecite{steinhoff_influence_2014}].
	Here, we follow a similar procedure and make use of the Wannier function continuum electrostatics approach (WFCE) \cite{rosner_wannier_2015} to include the screening effects of substrates, as described in the following.
	
	The bare interaction matrix $U_{\alpha \beta}(q)$ in the orbital basis $\alpha$, $\beta \in \mathrm{\{ d_{z^2}, d_{xy}, d_{x^2y^2} \} }$ of the Mo atoms is obtained for the freestanding undoped material via RPA calculations using the Spex and Fleur software codes \cite{friedrich_efficient_2010, FLEUR}.
	To parametrize the Coulomb interaction, we use the sorted eigenbasis of the bare interaction to diagonalize the latter 
		\begin{equation}
		\mathbf{U}^\mathrm{diag}(q) = \left(\begin{array}{ccc}
		U_1^\mathrm{diag}(q) & 0 & 0\\
		0 & U_2^\mathrm{diag} & 0 \\
		0 & 0 & U_3^\mathrm{diag}
		\end{array} \right).
		\label{eq:BareInteractionMatrixEigenBasis}
		\end{equation}
		Here, the diagonal matrix elements are given by
		\begin{equation}
		U_i^\mathrm{diag} = \langle e_i | \mathbf{U} | e_i \rangle
		\end{equation}
		using the eigenvectors of $\hat U(q \rightarrow 0)$
		\begin{equation}
			e_1=\begin{pmatrix}1/\sqrt{3}\\1/\sqrt{3}\\1/\sqrt{3}\end{pmatrix}, e_2=\begin{pmatrix}\sqrt{2/3}\\-1/\sqrt{6}\\-1/\sqrt{6}\end{pmatrix},
			e_3=\begin{pmatrix}0\\1/\sqrt{2}\\-1/\sqrt{2}\end{pmatrix}.
			\label{eq:Eigenvectors}
		\end{equation}
		$U_1^\mathrm{diag}(q)$ is the leading eigenvalue of the bare interaction and the other two eigenvalues are approximately constant.
	For the leading eigenvalue, we obtain a fit of the form
	\begin{equation}
	U_1^{\rm diag}(q) = \frac{3 e^2}{2 \varepsilon_0 A} \frac{1}{q(1 + \gamma q)}
	\label{eq:BareInteraction}
	\end{equation}
	with the area of the 2d hexagonal unit cell $A= \frac{\sqrt{3}}{2} a^2$ and the lattice parameter $a=$ 3.18\,\AA . 
	The factor $3$ in Eq. (\ref{eq:BareInteraction}) arises from the fact that we use three orbitals to describe the system and treat the Coulomb interaction in the eigenbasis of the bare interaction.
	$\gamma$ describes how the effective height affects short wavelengths, which means that it is a structure factor and becomes important at large wavevectors $q$.
	The value of $\gamma$ is given in Tab. \ref{tab:Fits}.
	
	\begin{table}[bt]
	\caption{Parameters describing the Coulomb interaction in \ce{MoS2}.}
	\label{tab:Fits}
	\begin{ruledtabular}
	\begin{tabular}{lrl}
    Parameter & Value &\\
    \hline
    $\gamma$ (\AA ) & 2.091 &\\
    $U_2^\mathrm{diag}$ (eV) & 0.810 &\\
    $U_3^\mathrm{diag}$ (eV) & 0.367 &\\
    $\varepsilon_\infty$ & 9.253 &\\
    $d$ (\AA ) & 9.136 &\\
    $\varepsilon_2$ & 3.077 &\\
    $\varepsilon_3$ & 2.509 &\\
	\end{tabular}
	\end{ruledtabular}
	\end{table}
	
	The screened matrix elements in the eigenbasis of the bare interaction are then obtained for the undoped system via 
	\begin{equation}
	V_i^\mathrm{diag}(q) = \left[ \varepsilon_i^\mathrm{diag}(q) \right]^{-1} U_i^\mathrm{diag}(q)
	\label{eq:BGScreenedInteraction}
	\end{equation}
	where $\varepsilon^{\rm diag}_i(q)$ accounts for the material specific interband polarizability and the polarizability of the substrate.
	
\begin{figure}[bt]
	\includegraphics[width=0.45\textwidth]{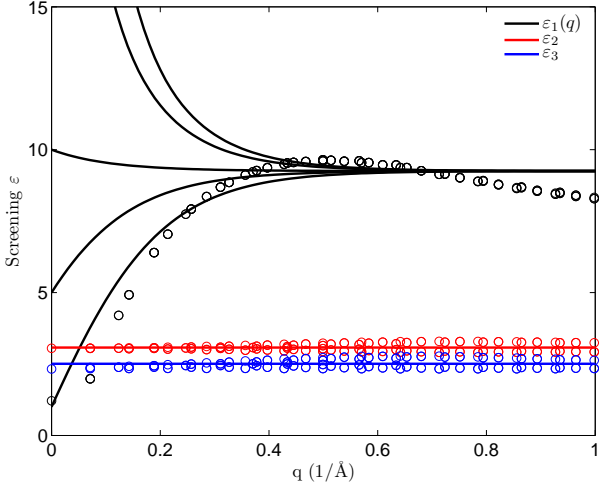}
	\caption{(Color online) Diagonal elements of the dielectric matrix in eigenbasis of the Coulomb interaction. $\epsilon_1$ belongs to the largest (macroscopic) eigenvalue of $U(q)$ while $\varepsilon_2$ and  $\varepsilon_3$ belong to the microscopic eigenvalues. From bottom to top, the substrate dielectric constants used for $\varepsilon_1(q)$  are $\varepsilon_\mathrm{sub}=\,$1, 5, 10, 50, $\infty$. The empty circles show the ab initio data for the screening without substrates.}
	\label{fig:Epsilon}
	\end{figure}

	Its diagonal represantation is given by
	\begin{equation}
	\varepsilon^\mathrm{diag}(q) = \left(\begin{array}{ccc}
	\varepsilon_1(q) & 0 & 0\\
	0 & \varepsilon_2 & 0 \\
	0 & 0 & \varepsilon_3 
	\end{array} \right)
	\end{equation}
	where the constants $\varepsilon_2$ and $\varepsilon_3$ (see  Tab. \ref{tab:Fits}) describe \emph{microscopic} screening effects which are similar to the bulk.
	The \emph{macroscopic} effects are described by the leading eigenvalue via
	\begin{equation}
	\varepsilon_1(q) = \varepsilon_\infty \frac{1 - \beta_1 \beta_2 e^{-2\,q\,d}}{1 + (\beta_1 + \beta_2) e^{-q\,d} + \beta_1 \beta_2 e^{-2\,q\,d}}
	\label{eq:EpsilonMacro}
	\end{equation}
	with
	\begin{equation}
	\beta_i	= \frac{\varepsilon_\infty - \varepsilon_\mathrm{sub, i}}{\varepsilon_\infty + \varepsilon_\mathrm{sub, i}}.
	\label{eq:BackgroundDielectricConstants}
	\end{equation}
	The involved parameters are derived from fits to the ab initio calculations (see Tab. \ref{tab:Fits}) for the freestanding layer. 
	The surrounding substrates have dielectric constants $\varepsilon_\mathrm{sub, 1}$ above and $\varepsilon_\mathrm{sub, 2}$ below the monolayer which can be varied using Eq. (\ref{eq:BackgroundDielectricConstants}).
	In the case of vacuum surrounding the monolayer ($\varepsilon_\mathrm{sub, 1}=\varepsilon_\mathrm{sub, 2}=1$) Eq. (\ref{eq:EpsilonMacro}) simplifies to
	\begin{equation}
	\label{eq:EpsilonFit}
	\varepsilon_1(q) = \varepsilon_\infty \frac{\varepsilon_\infty + 1 - (\varepsilon_\infty - 1) e^{-q\,d}}{\varepsilon_\infty + 1 + (\varepsilon_\infty - 1)  e^{-q\,d}}.
	\end{equation}
	The macroscopic screening for various substrates and the microscopic screening are shown in Fig. \ref{fig:Epsilon}.
	
	Once we have obtained the diagonal dielectric matrix $\varepsilon^{\rm diag}(q)$ we can calculate the screened Coulomb interaction in the eigenbasis using Eq. (\ref{eq:BGScreenedInteraction}) together with Eqs. (\ref{eq:BareInteractionMatrixEigenBasis}) and (\ref{eq:BareInteraction}) and the parameters in Tab. \ref{tab:Fits}.
	Afterwards, we can transform to the orbital basis using the eigenvectors in Eq. (\ref{eq:Eigenvectors}).
	This analytic description allows to evaluate the bare and screened Coulomb matrix elements at arbitrary momenta $q$ in the first Brillouin zone and for arbitrary dielectric environments.
	
	To get the real space values we need to do a simple Fourier transform, resulting in the onsite bare and screened Coulomb matrix elements given in Tab. {\ref{tab:UJVW}}.
	
	\begin{table*}[hbt]
		\caption{Bare onsite $U$ as well as background screened onsite $V$ and fully screened onsite Coulomb matrix elements $W$ for the three important orbitals in real space. Values for $W$ are in the range of low electron doping $x \approx 0.04$ \cite{DopingLevel} in the fifth ($W_\mathrm{low}$, K is occupied) and for high electron doping $x \approx 0.13$ in the last column ($W_\mathrm{high}$, K and $\Sigma$ are occupied).}
		\label{tab:UJVW}
		\begin{ruledtabular}
		\begin{tabular}{llcccc}
		 & & bare & undoped & \multicolumn{2}{c}{doped}\\
		\multicolumn{2}{c}{orbitals}&$U$ (eV)&$V$ (eV)&$W_\mathrm{low}$ (eV)&$W_\mathrm{high}$ (eV)\\
		\hline
		d$_\mathrm{{z^2}}$    & d$_\mathrm{{z^2}}$    & 9.11 & 1.55 & 0.82 & 0.68 \\
		d$_\mathrm{{z^2}}$    & d$_\mathrm{{xy}}$     & 8.30 & 1.29 & 0.58 & 0.44 \\
		d$_\mathrm{{z^2}}$    & d$_\mathrm{{x^2y^2}}$ & 8.30 & 1.29 & 0.58 & 0.44 \\
		d$_\mathrm{{xy}}$     & d$_\mathrm{{xy}}$     & 8.89 & 1.49 & 0.80 & 0.64 \\
		d$_\mathrm{{xy}}$     & d$_\mathrm{{x^2y^2}}$ & 8.52 & 1.35 & 0.65 & 0.51 \\
		d$_\mathrm{{x^2y^2}}$ & d$_\mathrm{{x^2y^2}}$ & 8.89 & 1.49 & 0.80 & 0.64 \\
		\end{tabular}
		\end{ruledtabular}
		\end{table*}

	\section{BCS equations in the multi-valley case}
	
	To discuss the possibility of unconventional Coulomb-driven superconductivity, we use the anisotropic BCS equations \cite{bennemann_physics_2003}
	\begin{equation}
	\Delta_k = -\frac{1}{N} \sum_{k'} W(k,k') \frac{\Delta_{k'}}{2 E(\varepsilon_{k'})} \tanh \frac{\beta E(\varepsilon_{k'})}{2},
	\label{eq:BCSAnisotropic}
	\end{equation}
	where $\Delta_k$ is the anistropic gap, $N$ is the number of unit cells, $W(k,k')$ is the coupling of momenta $k$ and $k'$, $E(\varepsilon_k) = \sqrt{\varepsilon_k^2 + \Delta_k^2}$ is the energy and $\beta$ is the inverse temperature.
	If the Fermi surface can be divided into different valleys $\mathrm{FS}_i$ [as it is the case in the doped TMDCs, cf. Fig. \ref{fig:bands_FS}(a)] and if the energy gap $\Delta$ is constant in each valley it only depends on the valley index
	\begin{equation}
	\Delta_k = \Delta_i \quad \mathrm{if} \, \,  k \in \mathrm{FS}_i.
	\end{equation}
	In this way, the summation over $k'$ in Eq. (\ref{eq:BCSAnisotropic}) can be decomposed into a summation over Fermi suface sheets and a summation over all momenta $k'$ on the Fermi surface sheet.
	If we further rewrite the summation over $k'$ into an energy integration and introduce the partial density of states per valley $N_i = \frac{1}{N} \sum_{k \in \mathrm{FS}_i} \delta(\varepsilon_{k'}-\varepsilon)$, we arrive at
	\begin{equation}
	\Delta_i = - \sum_{j} W_{ij} \Delta_{j} N_j 2 \int_{0}^{E_\mathrm{cut}} \mathrm{d \varepsilon} \frac{1}{2 E(\varepsilon)} \tanh \frac{\beta E(\varepsilon)}{2}
	\end{equation}
	assuming a constant coupling $W_{ij}$ inside the valleys and an energy cutoff $E_\mathrm{cut}$.
	
	Close to the critical temperature, the gap is $\Delta \approx 0$ and we can rewrite the integration to 
	\begin{equation}
	\int_{0}^{\beta E_\mathrm{cut} / 2} \mathrm{d}x \frac{\tanh x}{x} \equiv F(\beta E_\mathrm{cut} / 2)
	\end{equation}
	 with the function $F(x)$ and $x=\beta \varepsilon /2$.
	This leads to
	\begin{equation}
	\Delta_i = - \sum_{j} W_{ij} \Delta_{j} N_j F\left(\frac{\beta E_\mathrm{cut}}{2}\right).
	\label{eq:BCSValley}
	\end{equation}
	
	We define a dimensionless valley-valley coupling constant $\mu_{ij} = 1 / N(\EF) V_{ij} N_i N_j$, where $N(\EF)$ is the total density of states per spin and the full coupling matrix is given by $(\boldsymbol{\mu})_{ij} = \mu_{ij}$.
	Using this terminology, Eq. (\ref{eq:BCSValley}) becomes
	\begin{equation}
	\Delta_i = - \sum_{j} \frac{N(\EF)}{N_i} \mu_{ij}  F\left(\frac{\beta E_\mathrm{cut}}{2}\right) \Delta_{j}
	\end{equation}
	which can be cast into the matrix form
	\begin{equation}
	0 = \left( \mathbf{N} + N(\EF)\, F \, \boldsymbol{\mu} \right) \boldsymbol{\Delta},
	\end{equation}
	where $\mathbf{N}$ is the diagonal matrix containing the density of states per valley and $\boldsymbol{\Delta}$ is a vector of the gaps in each valley.
	
	A non-trivial solution exists if 
	\begin{equation}
	\det\left[\mathbf{N} + N(\EF)\, F \, \boldsymbol{\mu} \right] = 0,
	\end{equation}
	which is not possible if all eigenvalues of $\mathbf{N}$ and $\boldsymbol{\mu}$ are positive.
	Since $\mathbf{N}$ has only positive eigenvalues, the necessary criterion is that at least one eigenvalue of $\boldsymbol{\mu}$ is negative.
	A sufficient criterion for a non-trivial solution and a superconducting instability is a negative eigenvalue of $\mathbf{N}^{-1/2} \boldsymbol{\mu} \mathbf{N}^{-1/2}$.
	
	If $\boldsymbol{\mu}$ is a $2 \times 2$-matrix with only positive, i.e. repulsive entries [cf. Eq. (\ref{eq:muinterintra})], it has a negative eigenvalue once the off-diagonal element is larger than the diagonal elements, which means in our case that the coupling between two valleys ($\mu_\mathrm{inter}$) is larger than the coupling inside of one valley ($\mu_\mathrm{intra}$) \cite{roldan_interactions_2013}.
	If the Fermi surface consists of more than two valleys, as it is the case for high electron doping in \ce{MoS2} where $\boldsymbol{\mu}$ is an $8 \times 8$-matrix, there are various $\mu_\mathrm{inter}$ and $\mu_\mathrm{intra}$, and it is more insightful to discuss the eigenvalues of the $\boldsymbol{\mu}$-matrix.

\end{appendix}

\bibliography{bibliography}

\begin{thebibliography}{10}
\bibitem{DopingLevel}
The electron doping level is given as the number of additional electrons per
  unit cell of molybdenum disulfide.
\end{thebibliography}
\bibliographystyle{h-physrev}

\end{document}